\documentclass[sigconf]{acmart}

\usepackage{xcolor,colortbl}
\usepackage{xspace}
\usepackage{url}
\usepackage{graphicx}
\usepackage{subcaption}
\usepackage{mwe}

\AtBeginDocument{%
  \providecommand\BibTeX{{%
    \normalfont B\kern-0.5em{\scshape i\kern-0.25em b}\kern-0.8em\TeX}}}

\setcopyright{acmcopyright}
\copyrightyear{2021}
\acmYear{2021}
\acmDOI{10.1145/1122445.1122456}

\acmConference[ESEM '21]{ESEM '21: 15th ACM/IEEE International Symposium on Empirical Software Engineering and Measurement}{October 11--15, 2021}{Bari, Italy}
\acmBooktitle{ESEM '21: 15th ACM/IEEE International Symposium on Empirical Software Engineering and Measurement, October 11-15, 2021, Bari, Italy}
\acmPrice{15.00}
\acmISBN{978-1-4503-XXXX-X/18/06}

\newif\ifdraft
\drafttrue

\newcommand{\nb}[2]{
	{
		{\color{black}{
				\fbox{\bfseries\sffamily\scriptsize#1}
				{\sffamily$\triangleright~${\it\sffamily #2}$~\triangleleft$}
	}}}
}

\ifdraft
\newcommand\jgb[1]{\nb{Jesús}{\color{blue}#1}}
\newcommand\grex[1]{\nb{Gregorio}{\color{red}#1}}
\newcommand\david[1]{\nb{David}{\color{green}#1}}

\newcommand\reviewer[1]{\nb{Reviewer}{\color{red}#1}}
\newcommand{\fixme}[1]{{\textcolor{red}{[FIXME] #1}}\xspace}

\else
\usepackage[disable]{todonotes}
\newcommand\jgb[1]{}
\newcommand\grex[1]{}
\newcommand\david[1]{}
\newcommand\dizquierdo[1]{}
\newcommand\reviewer[1]{}
\newcommand{\fixme}[1]{}

\fi

\usepackage[inline]{enumitem}

\newcommand{\ie}{\emph{i.e.,}\xspace}
\newcommand{\eg}{\emph{e.g.,}\xspace}

\newcommand{\etal}{\emph{et~al.}\xspace}

\newcommand{\figref}[1]{Figure~\ref{#1}\xspace}

\newcommand{\tool}[1]{{\sc #1}\xspace}
\newcommand{\babia}{\tool{BabiaXR}}
\newcommand{\kibana}{\tool{Kibana}}
\newcommand{\elasticsearch}{\tool{ElasticSearch}}
\newcommand{\grimoirelab}{\tool{GrimoireLab}}

\newcommand{\babiarepo}{\url{https://gitlab.com/babiaxr/aframe-babia-components}}

\newcommand{\babianpm}{\url{https://npmjs.org/package/aframe-babia-components}}

\newcommand{\webxr}{\tool{WebXR}}
\newcommand{\webgl}{\tool{WebGL}}

\newcommand{\aframe}{\tool{A-Frame}}

\newcommand{\ack}{We acknowledge the financial support of the Spanish Government for the projects IND2018/TIC-9669 and RTI-2018-101963-B-I00.}

\begin{document}

\title{To VR or not to VR: Is virtual reality suitable to understand software development metrics?}

\author{David Moreno-Lumbreras}
\email{d.morenolu@alumnos.urjc.es}
\affiliation{%
  \institution{EID @ Universidad Rey Juan Carlos, Bitergia}
  \city{Móstoles}
  \country{Spain}
}

\author{Gregorio Robles}
\email{gregorio.robles@urjc.es}
\affiliation{%
  \institution{Universidad Rey Juan Carlos}
  \city{Madrid}
  \country{Spain}
}

\author{Daniel Izquierdo-Cort\'azar}
\email{dizquierdo@bitergia.com}
\affiliation{%
  \institution{Bitergia}
  \city{Madrid}
  \country{Spain}}

\author{Jesus M. Gonzalez-Barahona}
\email{jesus.gonzalez.barahona@urjc.es}
\affiliation{%
  \institution{Universidad Rey Juan Carlos}
  \city{Madrid}
  \country{Spain}
}

\renewcommand{\shortauthors}{Moreno-Lumbreras et al.}

\begin{abstract}

  {\bf Background/Context:}
  Currently, the usual interface for visualizing data is based on 2-D screens. Recently, devices capable of visualizing data while immersed in VR scenes are becoming common. However, it has not been studied in detail to which extent these devices are suitable for interacting with data visualizations in the specific case of data about software development.
\\
{\bf Objective/Aim:}
In this registered report, we propose to answer the following question: ``Is comprehension of software development processes, via the visualization of their metrics, better when presented in VR scenes than in 2D screens?'' In particular, we will study if answers obtained after interacting with visualizations presented as VR scenes are more or less correct than those obtained from traditional screens, and if it takes more or less time to produce those answers.
\\
    {\bf Method:}
  We will run an experiment with volunteer subjects from several backgrounds. We will have two setups: an on-screen application, and a VR scene. Both will be designed to be as much equivalent as possible in terms of the information they provide. For the former, we use a commercial-grade set of \kibana-based interactive dashboards that stakeholders currently use to get insights. For the latter, we use a set of visualizations similar to those in the on-screen case, prepared to provide the same set of data using the museum metaphor in a VR room. The field of analysis will be related to modern code review, in particular pull request activity. The subjects will try to answer some questions in both setups (some will work first in VR, some on-screen), which will be presented to them in random order. To draw results, we will compare and statistically analyze both the correctness of their answers, and the time spent until they are produced.

\end{abstract}

\ccsdesc[500]{Software and its engineering~Software development process management}
\ccsdesc[500]{Software and its engineering~Software development methods}

\keywords{virtual reality, dashboards, controlled experiment, code review, pull requests}

\maketitle

\section{Introduction}
\label{sec:intro}


Software engineers mainly interact with source code using a keyboard and a mouse, typically viewing it on 2D monitors.
This way of interacting does not take advantage of the many benefits of movement and perception that humans have.
This is not the case of virtual reality (VR) immersion, where the subject works in a virtual 3D environment. In the last years, affordable VR devices have emerged, and new standards, such as \webxr~\cite{jones20:webxr_devic_api} and \webgl~\cite{jackson20:webgl}, have become available, so that VR applications can be made portable to different platforms, and easily integrable with other applications and APIs. Therefore, we are now at a point where using VR for interacting with data visualizations is feasible in many environments.

In fact, in the specific case of software engineering, some scholars argue that the use of VR allows for environments that may make practitioners face lower learning curves, be more creative and achieve higher productivity~\cite{elliott2015virtual}. However, to our knowledge, there is little evidence that VR can provide better, or even similar results to on-screen visualizations, in the specific case of software development data.

Bitergia\footnote{\url{https://www.bitergia.com}} is a company offering commercial services in the area of software development analytics, specialized in offering software development data visualizations in web-based 2D dashboards.
Recently, to explore visualizing software development data in VR, Bitergia and the Universidad Rey Juan Carlos have developed \babia,\footnote{\url{https://babiaxr.gitlab.io/} (see also section~\ref{sec:tools})} a toolset for visualizing data in 3D, both on-screen and in VR devices.

The fact that we can design a visualization of exactly the same data, with very similar charts, in two different settings (2D charts in traditional screens, and 3D charts in VR devices) gives us the chance of trying fair comparisons between both. In particular, we can design VR scenes with charts similar to the 2D, on-screen visualizations that Bitergia uses for its customers, and then run an experiment where subjects use those settings so that we can compare results.

Hence, in this report we propose a controlled experiment for the comparative evaluation of two approaches: {\bf 2D data visualization \emph{vs.} VR immersive visualization} of data about software development.
The main aim is to test if VR immersion is at least as effective and efficient as on-screen 2D for the visualization of the same data. We will design a set of comprehension tasks, formulated as the answer to a question, and analyze both the correctness of answer and the time to get an answer.

The field of analysis for our experiment will be pull request activity. Pull request, as part of modern code review~\cite{bacchelli2013expectations,thongtanunam2017review}, is a software development activity that has been widely researched by academia in the last years~\cite{kononenko2018studying,maddila2019predicting,yu2015wait}.
It is of major interest to industry and practitioners as it is very human-intensive and often the cause of bottlenecks and inefficiencies~\cite{sadowski2018modern}.

\section{Background}

\subsection{2D Visualizations}
\label{sec:2dvis}

With the help of \grimoirelab,\footnote{\url{https://chaoss.github.io/grimoirelab/} (see also section~\ref{sec:tools})} Bitergia gathers and enriches software development data
to create a set of pre-existing dashboards. In particular, there are five panels/dashboards whose main
goal is to show the several aspects of the pull request process.

\begin{figure*}[ht]
  \centering
  \begin{subfigure}[b]{0.49\textwidth}
    \centering
    \includegraphics[width=\textwidth]{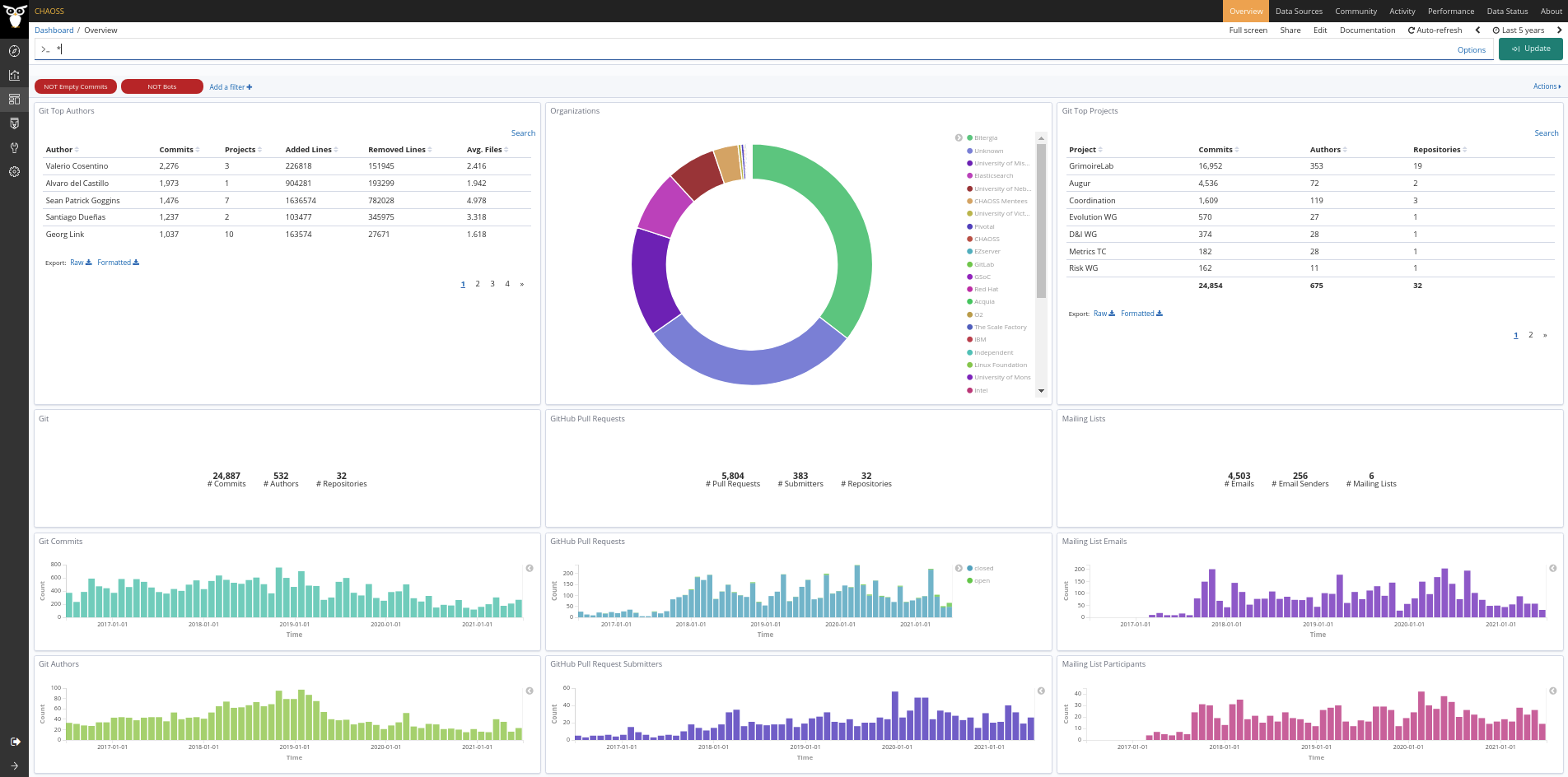}
    \caption[Overview panel]%
    {{\small Overview}}
    \label{fig:overviewpanel}
  \end{subfigure}
  \hfill
  \begin{subfigure}[b]{0.49\textwidth}
    \centering
    \includegraphics[width=\textwidth]{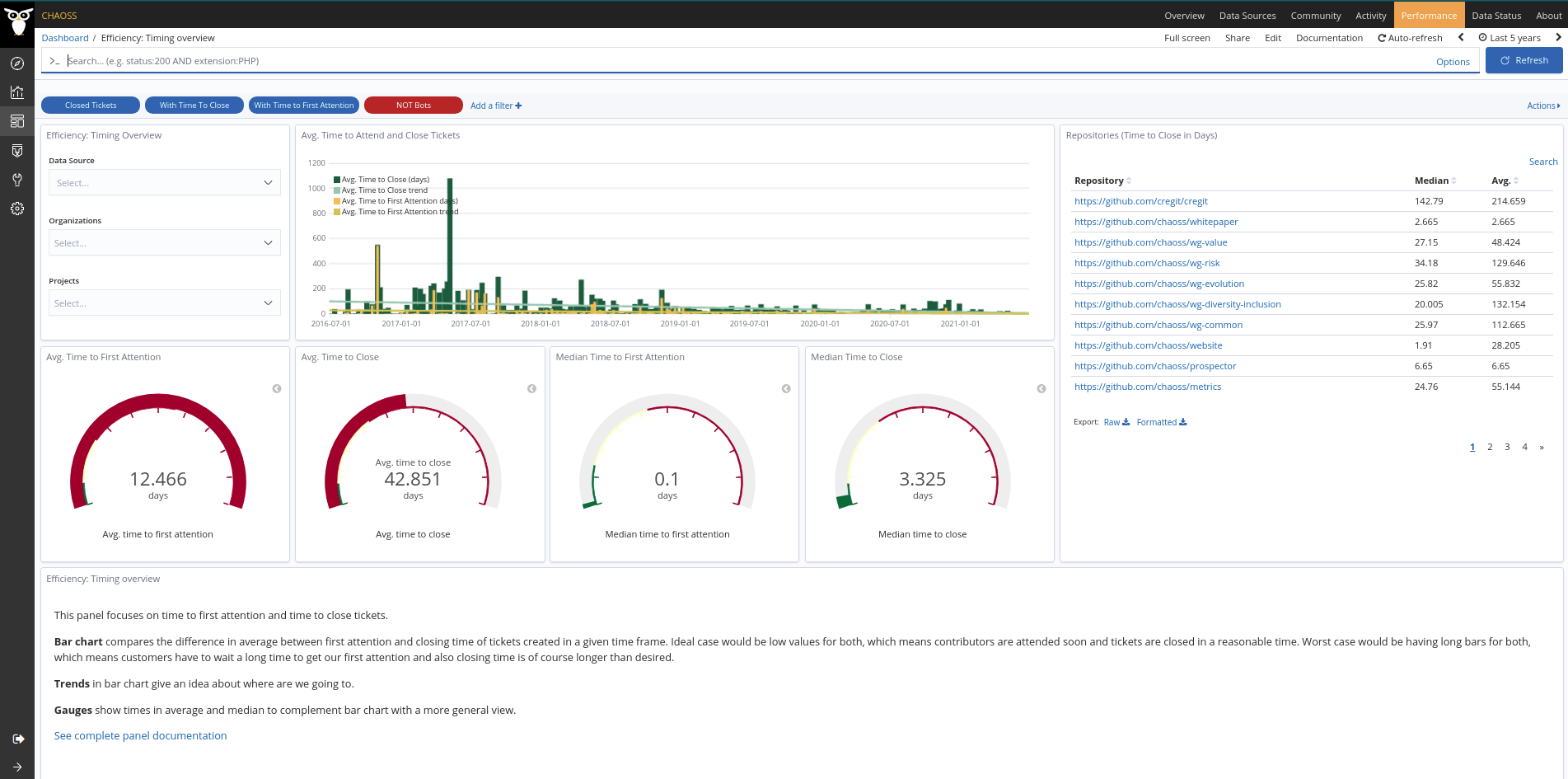}
    \caption[Efficiency panel]%
    {{\small Efficiency}}
    \label{fig:efficiencypanel}
  \end{subfigure}
  \vskip\baselineskip
  \begin{subfigure}[b]{0.49\textwidth}
    \centering
    \includegraphics[width=\textwidth]{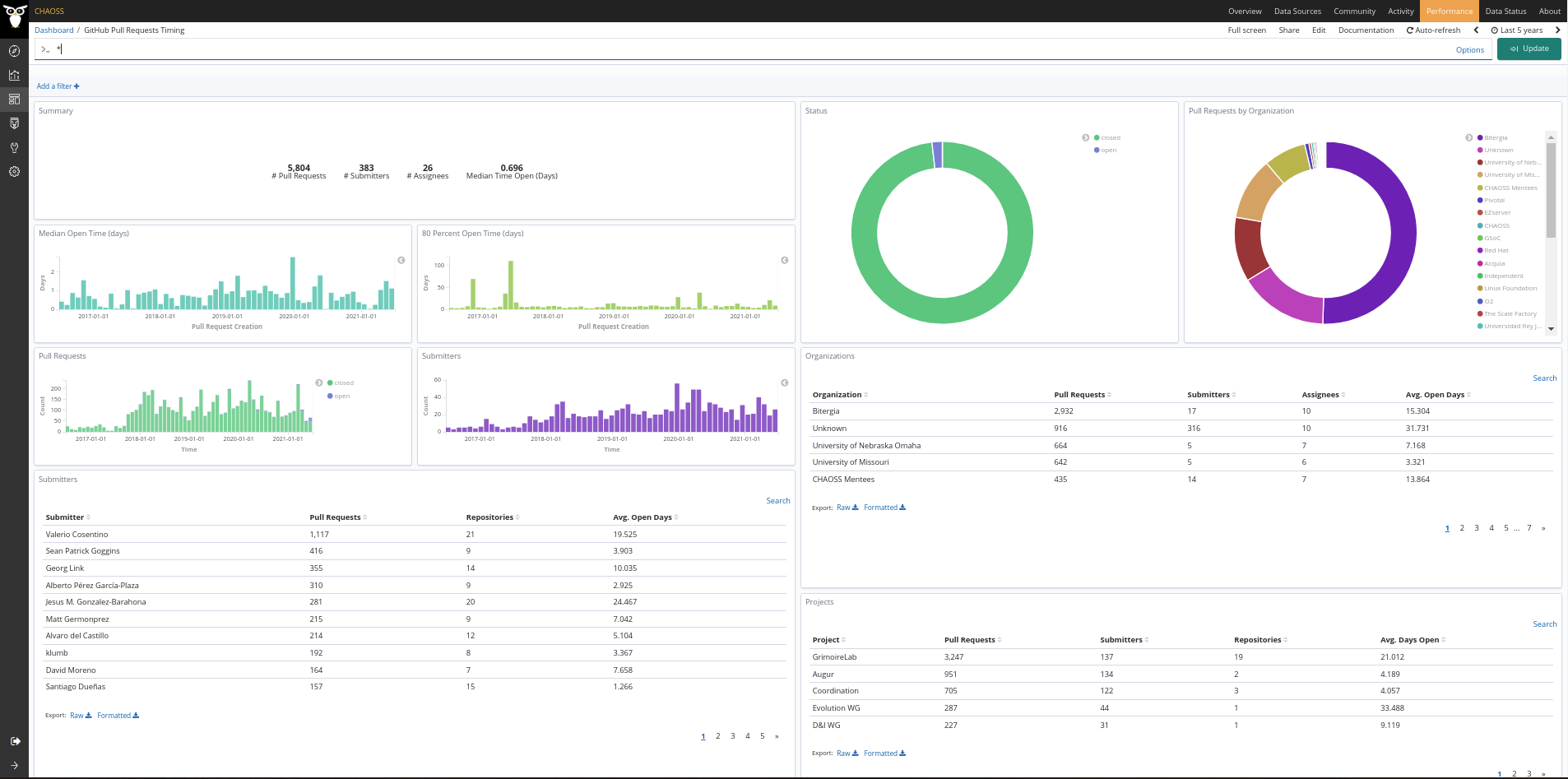}
    \caption[]%
    {{\small Timing}}
    \label{fig:timingpanel}
  \end{subfigure}
  \hfill
  \begin{subfigure}[b]{0.49\textwidth}
    \centering
    \includegraphics[width=\textwidth]{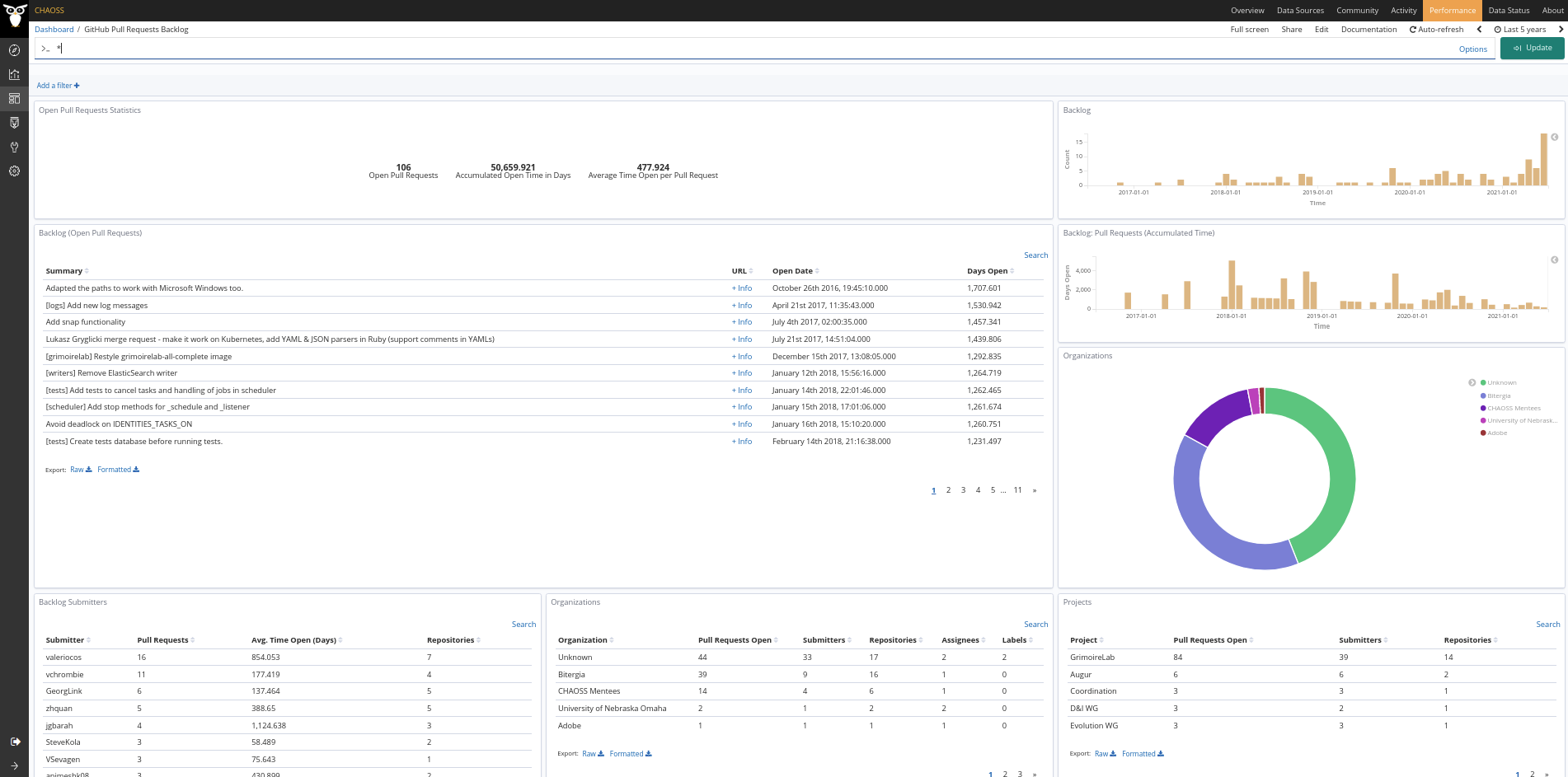}
    \caption[]%
    {{\small Backlog}}
    \label{fig:backlogpanel}
  \end{subfigure}
  \caption{\grimoirelab panels in \kibana}
  \label{fig:grimoirelabpanels}
\end{figure*}

\begin{enumerate}
  \item Overview (see~\figref{fig:overviewpanel}): provides a general overview of the activity and the authors
        of each of the pull requests. People and organizations are the key pieces of
        information to display.
  \item Efficiency (see~\figref{fig:efficiencypanel}): gives information about the general trend of the
        code review process and its time to merge pull requests. The main visualizations
        are the speedometer that helps understand how good the process has been in the
        chosen period of time. And two extra visualizations that provide the
        BMI (Backlog Management Index) for pull requests, a known
        metric in the maintenance ecosystem~\cite{malanga2015assessing}, and the lead time.
  \item Timing (see~\figref{fig:timingpanel}): displays information about the time to close pull requests. 
        It offers insight to the user about the evolution of the time
        to close a pull request over time. This is intended to visually find 
        bottlenecks in the development process.
  \item Backlog (see~\figref{fig:backlogpanel}): displays the remaining open pull requests over time and the newest
        and oldest of them. This aims at identifying old pull requests still open with
        the goal of cleaning this backlog and focus on the important ones.
  \item Time to Attention: in any development process the time until a project member
        answers a pull request is key to have a proper submitter-reviewer interaction. This
        panel helps understand how good the project is when collaborating with others
        or within the project.
\end{enumerate}

An example of the described panels in \kibana is shown in \figref{fig:grimoirelabpanels}. The data represented is not the data that will be used in the experiment, but the visualizations on them are the ones that will be used. All the panels represent the data with a set of different charts, from simple charts showing raw data as tables to other common charts as simple bars, pies, or line charts.

All these panels are now part of existing production solutions for several
organizations that are using them to analyze their code review process.
The CHAOSS Community or the FINOS Foundation are a couple of existing
and public dashboards\footnote{http://chaoss.biterg.io/} although these panels are deployed across all of the
customers of the Bitergia Analytics Platform that use GitHub. This includes
those using them in open source projects, or those using GitHub Enterprise
edition and analyzing internal developments.

These five panels are an example of the out-of-the-box functionality provided by
the \grimoirelab open source project. However, customers require certain
actions to polish the dashboards and adequate them to their needs.
The issue is indeed related to the lack of space to represent all of the data
that interests a stakeholder. The five panels cover specific use cases that have
become standard in the tool, but each of them needs the other four to
offer a complete picture of the code review process. This requires the data consumer
to move from one to the other, apply and fix filters across dashboards
and other actions to successfully navigate through them.

When designing the panels, the developers had a list of associated questions (i.e., goals) for each of them.
In particular:

\begin{enumerate}

  \item Overview: Number of organizations sending pull requests, status over time of the number of open and closed pull requests, number of people submitting pull requests over time, distribution of the organizations sending pull requests over time.

  \item Efficiency: How good is the project dealing with pull requests, the workload adequacy of the project and its evolution (Backlog Management Index - Review Efficiency Index), and lead time (latency between the initiation and completion of a process).

  \item Timing: Past bottlenecks, time to merge pull requests per organization.

  \item Backlog: Accumulated time of the backlog over time, backlog over time, oldest pull requests.

  \item Time to Attention: How fast are developers answering to a pull request.

\end{enumerate}

On the other hand, and as noted before, there are aspects that require to combine information in more than one panel.
Among these, we can list the following:

\begin{itemize}
  \item Organizational performance (number of still open pull requests, time to close a pull requests, time to be closed a pull request, lead time).

  \item Project performance (number of still open pull requests, time to close a pull requests, time to be closed a pull request, lead time).

  \item Relation between the time to answer a pull requests and the time to be reviewed or the time to review.

  \item Comparison of organizations contributing to a project by their pull request review time, existing open backlog (what they have opened).

\end{itemize}

\subsection{VR Visualizations}
\label{sec:vr-vis}

The goal of the VR environment is to take advantage of the 3 dimensions that it provides. We will use \babia for the development of VR visualizations, providing a scene where the visualizations will be organized.

\babia provides a set of different 3D visualizations, including some with three axis (e.g., bars, cylinders, bubbles) to represent depth, height, and width in the same visualization, and cylinder charts that offer the radius for expressing one more metric (compared to a common bars chart). \babia also provides 3D treemap visualizations that can be used to represent the code city metaphor.

\begin{figure}[ht]
  \centering
  \includegraphics[width=\columnwidth, keepaspectratio]{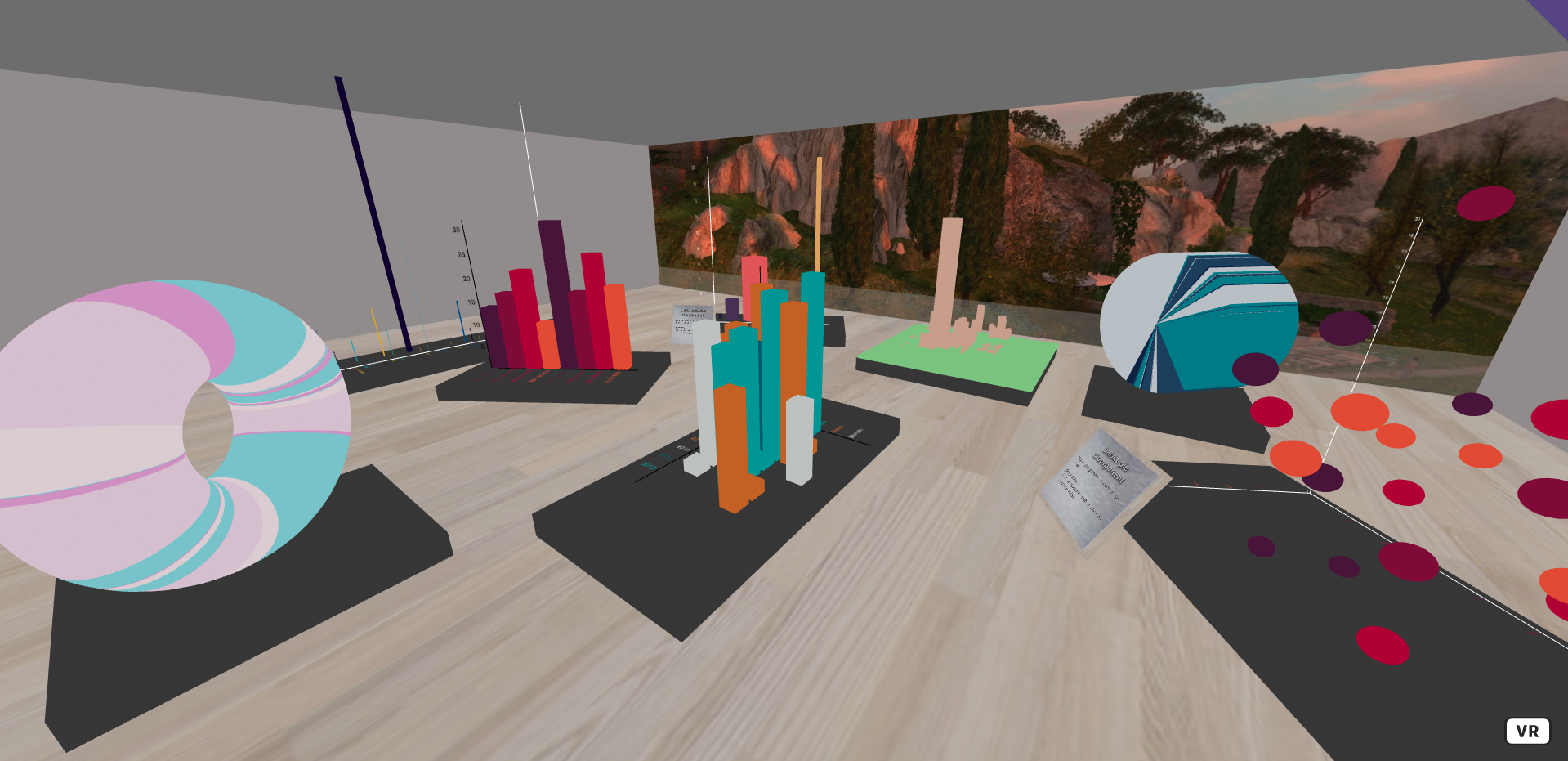}
  \caption{Example of a \babia scene}
  \label{fig:vis_babia}
\end{figure}

\figref{fig:vis_babia} shows an example of the scenes that can be created with \babia. The scene represented in the figure is not the scene that we will use in the experiment, although the general effect will be similar. In the scene we can see some visualizations that go beyond what \kibana can provide, taking advantage of the three dimensions. Thanks to them, we can aggregate two or more \kibana visualizations in one: for example, we can use the 3D bars map in \babia aggregating two or more simple \kibana barchart visualizations.

As already reported, the aim of this registered report is to represent the information displayed in the five panels described in Section~\ref{sec:2dvis} in a VR scene, having similar visualizations but taking advantage of the availability of new geometry metrics (e.g., new axis, radius, city) for combining or improving the presented in the on-screen environment.

\subsection{Related studies}

Wettel \etal~\cite{ref_survey} proposed one of the first experiments to validate the 3D visualization based on the city metaphor as a way to comprehend some aspects of software systems. Our experiment can have this type of visualization in the VR immersion, as \babia has included it in its stack. More recently, Romano \etal~\cite{ref_exp2} conducted a controlled experiment where they asked participants to perform program comprehension tasks with the support of the \tool{Eclipse} Integrated Development Environment with a plugin for gathering code metrics and identifying bad smells, and a visualization tool of the city metaphor displayed on a standard computer screen and in immersive virtual reality.
The results show that the participants using the VR environment were faster than the participants using the computer screen, but limited to the city metaphor. Our study will encompass more types of data visualizations. Merino \etal~\cite{MerinoExperiment} used the city metaphor visualization to visualize software systems in three settings: i) a computer screen, ii) an immersive 3D scene (VR), and iii) a physical 3D printed model. Then, they carried out an experiment that measured the effectiveness of visualizations in terms of performance, recollection, and user experience. They found that participants using the VR scene obtained the highest recollection, while participants using the 3D printed model were the fastest ones resolving tasks. Compared to this study, our experiment includes more types of visualization and will encompass more metrics and aspects about software development projects.

\section{Experiment planning}
\label{sec:experiment planning}

\subsection{Goal}

The main objective of the experiment is to analyze if the comprehension of software development processes (in particular, modern code review), via the visualization of their metrics, is better when presented in VR scenes than in 2D screens.

\subsection{Participants and settings}

Our experiment will be run with at least 30 subjects from both academia (students and researchers) and industry. All subjects will be screened via a questionnaire, to ensure a minimum knowledge about software development projects (only those passing the questionnaire will be considered for the experiment). 
Recruited subjects will not have any relation with the study or the paper. Before participating in the experiment, subjects will fill a demographics form, which will be used to determine the confounding variables.

Subjects will run the experiment in two settings, their order being randomly decided for each of them:

\textbf{On-screen}: The setting will be a computer with a web browser showing a \kibana dashboard, with several panels including all visualizations needed to answer the questions (as described in Section~\ref{sec:2dvis}).

\begin{figure}[ht]

  \centering
  \includegraphics[width=0.9\columnwidth, keepaspectratio]{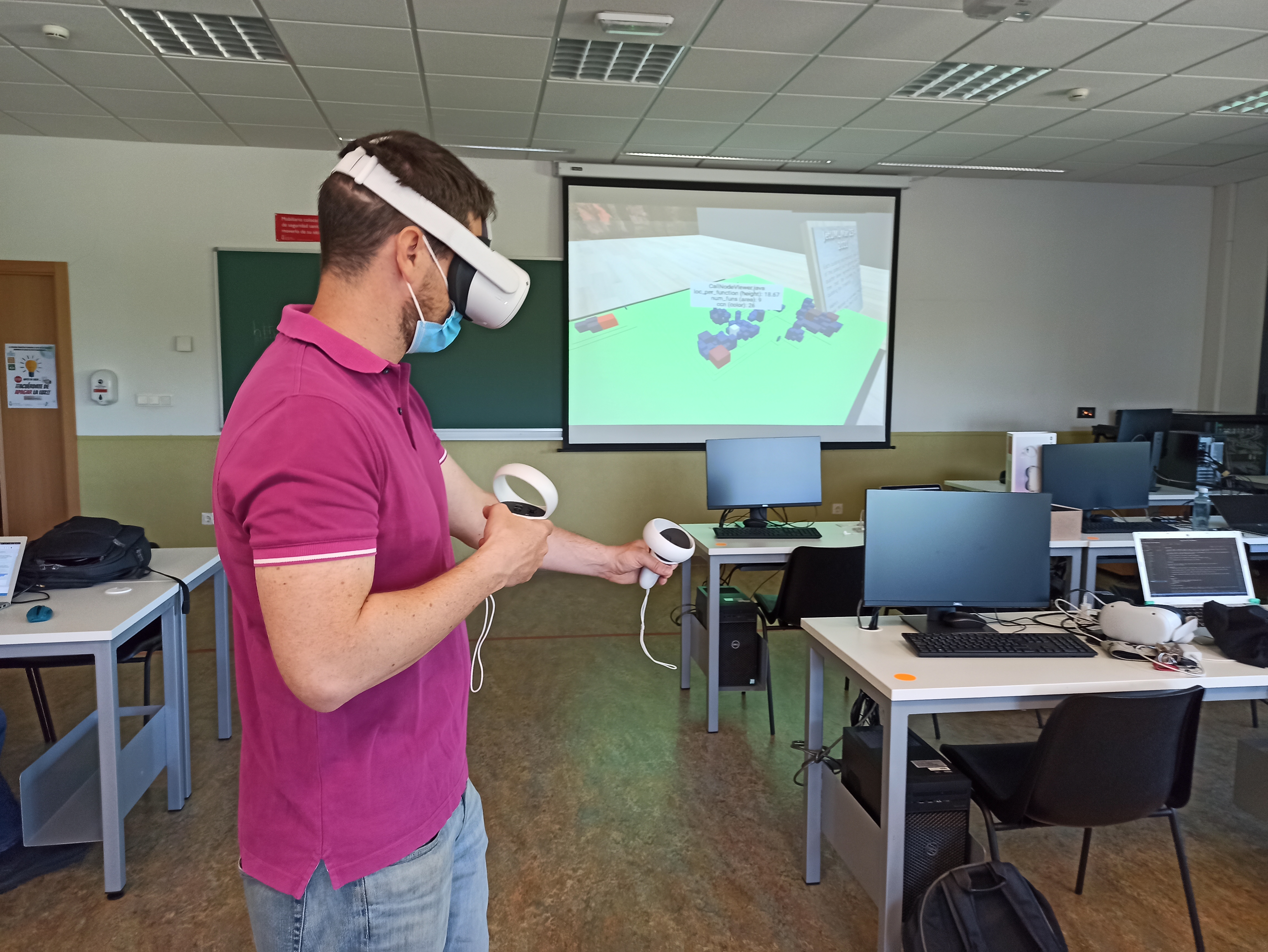}
  \caption{An example of the VR experiment setup}
  \label{fig:vr-experiment}
\end{figure}

\textbf{Virtual Reality}: The setting will be VR scene (as described in Section~\ref{sec:vr-vis}), in which subjects will immerse using the VR browser in an Oculus Quest headset.
\figref{fig:vr-experiment} shows a participant during the VR experiment and the screencast for the supervisor.

In each setting, subjects will be presented with visualizations for data from a different project (of the two projects that will be used), and the order in which projects are presented will also be random. All subjects will be presented with the same questions, but each question will be presented to each subject randomly only in one of the two settings, and for one of the two projects.

In both settings, subject's answers will be aloud, and the screen or VR scene as presented to the subject will be screencasted. Both the screencast and subject's answers will be recorded (and then transcribed). The setting will be followed by a supervisor, who provides support if needed, having access to the screencast and communicating with the subject via voice. If sanitary conditions recommend isolated work, the supervisor will participate remotely, following the screencast, and using some conferencing application to talk with the subject.

We will ensure that all the participants have experience in different relevant topics about software development using a questionnaire, reducing the threat that they are not competent enough.
We will also ask for their experience with VR and \kibana to mitigate the threat that the participant's experience is not fairly distributed.

\subsection{Tools}
\label{sec:tools}

For the data gathering of the project, we will use the \grimoirelab platform. \grimoirelab, a free, open source software for software development analytics~\cite{duenas21:grimoirelab}, is a toolset that includes modules to retrieve data from many kinds of software repositories~\cite{duenas2018perceval}, store it in a database, and then process and analyze it, producing many different metrics. It also includes visualization modules that can be used to interact with the data via traditional, on-screen, web browsers. \grimoirelab is now a project under the umbrella of the Linux Foundation CHAOSS community.\footnote{\url{https://chaoss.community}} 


Data produced by \grimoirelab can be fed to \babia as well, so that the same pipeline can be used to visualize data in 2D screens and in VR devices. \babia is a toolset, developed by the authors, for 3D data visualization in the browser. It is based on \aframe,\footnote{\aframe: \url{https://aframe.io}} an open web framework to build 3D, augmented reality (\ie AR), and VR experiences in the browser. \aframe extends \texttt{HTML} with new entities allowing to build 3D scenes as if they were \texttt{HTML} documents, using techniques common to any front-end web developer. \aframe is built on top of \tool{Three.js},\footnote{\tool{Three.js}: \url{https://threejs.org}} which uses the \webgl API available in all modern browsers.

\babia extends \aframe by providing components to create visualizations, simplify data retrieval, and manage data (\eg data filtering or mapping of fields to visualization features). Scenes built with \babia can be displayed on-screen, or in VR devices, including consumer-grade headsets. \figref{fig:vis_babia} shows a sample scene built with \babia. \babia is open source: Its source code is available on \tool{GitLab}\footnote{\babiarepo} and it can be installed with \tool{npm}.\footnote{\babianpm}

For the development of the on-screen participants, we will use \kibana. \kibana is a free and open frontend application that sits on top of the Elastic Stack, providing search and data visualization capabilities for data indexed in \elasticsearch. More details about the panels of \kibana are described in Section~\ref{sec:2dvis}.


\subsection{Datasets}
\label{sec:datasetssub}

The datasets used in the experiment will consist of data from two software development projects, to be defined. The characteristic of the projects will be similar for doing a fair comparison and decrease the threat of having an unbalanced amount of data in one of the settings. For the data retrieval, we will use \grimoirelab that provides a complete set of metrics of a software development project, focusing the experiment on the process analysis (efficiency of tickets/issues, merge/pull requests, etc.).



\subsection{Experimental material}
\label{sec:material}

To carry out the experiment we will use common computers for the on-screen part of the experiment. Since \kibana is a web application and can be used on any computer with a modern web browser, participants will be given the possibility to use their own -- but in the event that they do not have one, the experimenter will offer one. For the virtual reality phase, we will use Oculus Quest 2 VR glasses. The experimenter will offer the participant these glasses to carry out the experiment, but if the participant has glasses of the same model, they can use them. In any case, the scene shown in VR is a web scene based on webXR standards, so we plan a version so that any virtual reality device with a modern browser will have access to it, making the realization of this experiment possible in a remote setting, if required.

\subsection{Tasks}

With the help of Bitergia and its experience in analysis of software development projects, we will define a set of tasks to analyze both the correctness of the task solutions and the task completion time. These tasks are yet to be defined; they will be related to a general view of the pull request activity, to authors, to project efficiency, among others. 

\subsection{Hypotheses or research questions}
\label{sec:hypotheses}

The main research question of our study is:

\begin{quote}
  RQ: {\em ``Is comprehension of software development processes, via the visualization of their metrics, better when presented in VR scenes than in 2D screens?''}.
\end{quote}


This question tests the hypothesis that presenting visualizations in VR, where available real state is much more abundant (you can have visualizations all around you, up and down if needed), will allow for a better and faster understanding. The hypothesis is disputable because there are factors that work against it, such the difficulties that perspective and distance may be cause to the adequate perception of magnitudes.


For answering the research question, and validating (or not) the hypothesis, we will focus on specific, measurable aspects of the answers produced by the subjects: accuracy (as a proxy for correctness), and time to answer (as a proxy for efficiency). Thus, we can refine our main RQ in two:

\begin{quote}
  RQ1: {\em ``Answers are more accurate in VR than on-screen?''} (alternate: more accurate on-screen)
\end{quote}

\begin{quote}
  RQ2: {\em ``Answers are obtained faster in VR than on-screen?''} (alternate: faster on-screen)
\end{quote}

To ensure that questions to subjects are useful for answering these two RQs, we will design them so that their answers are numeric.

\subsection{Variables}
\label{sec:variables}





The independent variable that will be input for our experiment is the setting (on-screen dashboards in VR scene) in which each answer will be answered. Dependent variables will be the answers that our subjects produce, and more concretely, how much deviated they are from the correct answer, and how long it took to produce them.

In more detail, for each Experiment Question (EQ), the variables will be:

\begin{itemize}
  \item Independent variable:
        \begin{description}
          \item[Name]: Setting
          \item[Description]: Setting in which the question was answered by the subject
          \item[Scale]: Categorical: ``Screen'' or``VR''.
          \item[Operationalization]: Subject answers after interacting with visualizations in a 2-D screen, or after interacting with visualizations immerse in virtual reality.
        \end{description}
  \item Dependent variable ``time-to-answer'':
        \begin{description}
          \item[Name]: TimeToAnswer
          \item[Description]: Time to answer the question
          \item[Scale]: Real number
          \item[Operationalization]: Number of seconds from the question is understood to the answer is produced.
        \end{description}
  \item Dependent variable ``Error'':
        \begin{description}
          \item[Name]: Error
          \item[Description]: Error in answering the question, with respect to the right answer.
          \item[Scale]: Real number
          \item[Operationalization]: Relative error of the magnitude produced as answer with respect to the true value of that magnitude.
        \end{description}
  \item Confounding variable ``Experience with \kibana'':
        \begin{description}
          \item[Name]: ExperienceKibana
          \item[Description]: Overall experience with \kibana dashboards
          \item[Scale]: Categorical: ``None'', ``small'', ``medium'', ``high''
          \item[Operationalization]: Question to the subject, with the operations of the scale, including an indicative number of hours for each category.
        \end{description}
  \item Confounding variable ``Experience with VR'':
        \begin{description}
          \item[Name]: ExperienceVR
          \item[Description]: Overall experience with VR devices
          \item[Scale]: Categorical: ``None'', ``small'', ``medium'', ``high''
          \item[Operationalization]: Question to the subject, with the operations of the scale, including an indicative number of hours for each category.
        \end{description}
  \item Confounding variable ``Experience in software development'':
        \begin{description}
          \item[Name]: Experience
          \item[Description]: Overall experience in software development
          \item[Scale]: Integer number
          \item[Operationalization]: Question to the subject about their experience in software development, with a number of years of experience as answer.
        \end{description}
  \item Confounding variable ``Position'':
        \begin{description}
          \item[Name]: Position
          \item[Description]: Position of the subject.
          \item[Scale]: Categorical: ``practitioner'', ``academic'', ``student''
          \item[Operationalization]: Question to the subject about their position with the three options in the scale as answer.
        \end{description}

\end{itemize}

We will control for confounding variables by splitting the answers according to them, and applying statistical controls to distill their effects.

\subsection{Design}

To understand whether VR is well suited to present software development-related visualizations compared to the traditional 2D on-screen representation we will conduct a controlled experiment.
Therefore, we will follow the \emph{ACM SIGSOFT Empirical Standards}~\cite{acm_standard} for the experiment design, specifically those aspects that are relevant for the quantitative method for experiments with humans, to satisfy all essential attributes and a part of the desirable ones.
We will follow the design of a ''One Factor, Two Level'' experiment, being the factor the independent variable (the setting), and the two levels the two values it may have (VR or in-screen). Since participants will be presented randomly with the order for the settings (first VR or first on-screen), this experiment can be treated and formally divided into two ``One Factor, Two Level'' sub-experiments.

\subsection{Procedure}

The analysis will be completed with a qualitative study, based on (possibly semi-structured) interviews to a sample of the people who will go through the experiment, to learn about the details of their experience.

For preparing the study, we will follow some procedures such as preparing the settings to perform the experiment, and testing them in advance, to avoid some mistakes that only the actual testing of the experiments will show.

\subsection{Analysis procedure}

For determining the behavior we will use statistical analysis for the completion time of the tasks and we will measure the significant differences with the best metric according to the sample size of the participants. We will also collect feedback about the difficulty of the tasks, avoiding the deviations between times, and deleting the outliers. Since the number of participants could be as low as of 30 participants, estimating the distribution of answers seems difficult. Therefore, we will use non-parametric tests, which do not assume specific distributions, to assess if the differences between answers in both settings are statistically significant, and determine if the results are meaningful. In principle, we plan to run the \textit{Mann-Whitney U} Test~\cite{nachar} and calculate the corresponding \textit{Cliff's Delta effect size}~\cite{cliff}.

\section{Execution Plan}
\label{sec:execution}

Before conducting the experiment, to avoid bias due to the previous experience with on-screen or VR settings as much as possible, all subjects will go through a short training procedure, to make sure they understand how the \kibana-based dashboards, and the VR immersive scenes work. This is necessary because we expect many subjects to not have previous exposure to \kibana, nor to VR immersion.

Upon completion of the training, a demographic form will be used to control the subject confounding variables. Once the form is completed, the experiment will begin by showing the subject data from one of the two projects considered (selected randomly) in a randomly chosen setting (VR or on-screen). Each subject will be presented with a number of questions (half of the total number of questions, selected randomly), in sequence, with each new question being presented only after the previous one was answered. Then, each subject will follow the same procedure in the other setting, with data from the other project.

To end the experiment, the subject will answer a feedback form which will be analyzed for possible improvements or problems that the participant has encountered during the experiment.

Once the experiment has been carried out, we will analyze the possible deviations that have occurred in its development

\section{Threats to validity}
\label{sec:theatstovalidity}

\subsection{Internal validity}

Internal validity is related to uncontrolled factors that can influence the effectiveness. In our case it pertains to:

\begin{itemize}

  \item \textbf{Subjects.} We ensure that all participants have experience in different relevant topics about software development projects using a questionnaire, thus reducing the threat that they were not competent enough. We will also ask for their experience in VR and \kibana, so that we can later do the corresponding analysis to mitigate the threat of the influence of previous experience with them.

  \item \textbf{Tasks.} The choice of tasks (questions) could be biased in favor of one of the settings. We mitigate this threat by developing scenes that are valid for both VR and on-screen, with exactly the same tasks, so that the level of difficulty is as similar as possible. We also will include tasks that put both modes at a disadvantage: Tasks focused on precision could be easier on-screen, while tasks focused on locality could be easier in VR. Not controlled aspects (\eg the relative size of the buildings in VR) could have an influence on the results.

  \item \textbf{Training.} In both settings we will present a basic guide on how to use the settings for performing the tasks, we will offer texts about how the tool is used and how the interaction with the elements works. We also will investigate whether a practical tutorial on how to interact with a VR headset could reduce the experience gap between VR and on-screen, improving the correctness of the VR tasks.

\end{itemize}

\subsection{External validity}

External validity relates to the generalizability of the results of the experiment. In our case it pertains to:

\begin{itemize}

  \item \textbf{Sample Size.} The number of participants in the experiment can be relatively small. A larger sample would be needed to have more conclusive results. In order to determine if the results are statistically meaningful, we will run tests depending on this sample size.

  \item \textbf{Subjects.} We will mitigate the threat of subject representativeness by categorizing them, including their professinal position and the years of experience in the main software development projects topics, obtaining a balanced mix of academics and professionals.

  \item \textbf{Target Systems.} Another threat is represented by the choice of the target systems. We will ensure that the two systems presented will have similar characteristics in order to mitigate the threat of having unbalanced systems that can interfere with the accuracy (correctness) and efficiency (completion time) of the responses.

  \item \textbf{Experimenter Effect.} The experimenters can be the authors of the research, which may influence any subjective aspect of the experiment. For example, task solutions cannot be graded correctly. To mitigate this threat, another author can carry out part of the experiments as a supervisor. Both experimenters will build a model of the responses based on previous experiments in the literature. Even if we will try to mitigate this threat extensively, we cannot exclude all possible influences on the results of the experiment.

  \item \textbf{Time Measurement.} To mitigate the threat that task completion times are not affected by external factors, subjects will be required to answer via voice in both settings.
  

\end{itemize}

\section{Contributions and implications}
\label{sec:contributions}

There is little evidence about the convenience of using VR for visualizing and interacting with data, and in particular a relative lack of evidence comparing on-screen with VR immersion in software engineering. As VR devices become increasingly commonplace, they will be more used in software engineering, and it will be important to know as much as possible about how they impact practitioners, in comparison with traditional on-screen environments.

Our main contribution would be to offer the results of a controlled experiment where 2D and VR are compared in terms of accuracy and speed of the same, high-level software comprehension tasks.
It will allow to identify affordances, applications, and challenges of both approaches, but especially for the VR one as it has been underinvestigated --and is thus less known-- by the scientific community.
We will also provide practitioners --in particular, but not limited to, the software analytics company that drives the development of \babia-- with future directions, showing what works and what requires further development.
Finally, all software used in the experiment is available under a free/open source license, making it possible for others (researchers and practitioners) to inspect how we have addressed the problems, to reuse it at their convenience or to adapt it to their needs.

We envision, given the technological maturity that VR is achieving, that VR is going to gain momentum in many areas.
Software engineering should not be an exception to this.
Regardless of whether the results obtained in our controlled experiment are positive or negative, we foresee that other researchers can devote their efforts to further advance this field of knowledge.
We will use VR in our experiment with code reviews and from a more software process perspective.
Future research could focus on other software development activities (e.g., modeling, developing, testing, profiling) and with other perspectives in mind (e.g., source code, diagrams).


\begin{acks}
  \ack
\end{acks}

\bibliographystyle{ACM-Reference-Format}
\bibliography{./paper}

\end{document}